\documentstyle[a4,12pt]{article}
\begin{document}
\setlength{\baselineskip}{20pt}
\setlength{\parskip}{10pt}

\null \ 
\vfill 
{\LARGE Uncertainty principle for proper time and mass}

\vskip 8mm
{
\setlength{\leftskip}{1cm}
\noindent
Shoju Kudaka \\
{\it Department of Physics, University of the Ryukyus, Okinawa, Japan}\\
Shuichi Matsumoto\footnote{Electronic mail address : shuichi@edu.u-ryukyu.ac.jp} \\
{\it Department of Mathematics, University of the Ryukyus, Okinawa, Japan}
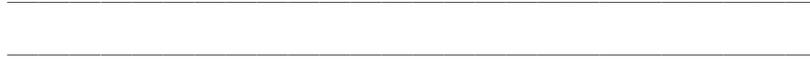
\vfill
\noindent
------------------------------------------------------------------------------\hfill \break 
------------------------------------------------------------------------------\hfill \break
\vfill
We review Bohr's reasoning in the Bohr-Einstein debate on the photon box experiment. The essential point of his reasoning leads us to 
an uncertainty relation between the proper time and the rest mass of the clock. It is shown that this uncertainty relation can be derived if only we take the fundamental point of view that the proper time should be included as a dynamic variable in the Lagrangian describing the system of the clock. Some problems and some positive aspects of our approach are then discussed.  PACS numbers: 03.65.Bz, 03.20.+i, 04.20.Cv, 04.60.Ds.
}
\vfill
\eject

\noindent
{\bf I. INTRODUCTION}

In various arguments about time, perhaps the most spectacular is the Einstein-Bohr debate on the photon box experiment$^{1,2}$. Their concern in the debate was Heisenberg's time-energy uncertainty relation. However, Bohr's reasoning reveals, as shown in the following, an uncertainty relation between the proper time and the rest mass of a clock. In fact, his essential point was simply that the very act of weighing a clock, according to general relativity, interferes with the rate of the clock.

In order to review Bohr's reasoning, we consider an experiment in which we measure the rest mass of a clock. We assume, of course, that the clock keeps its own proper time.      

Following Einstein's stratagem, we try to weigh the clock by suspending it with a spring. That is to say, if the spring stretches by the length $l$, we can calculate the mass $m$ of the clock from the relation 
$$kl=mg,$$
where $g$ is the gravitational acceleration and $k$ is a constant characterizing the spring. 

Assume that a scale is fixed to the spring support, and that we read the length $l$ on it with an accuracy $\Delta q$. Then the determination of $l$ involves a minimum latitude $\Delta p$ in the momentum of the clock, related to $\Delta q$ by the equation $\Delta q \Delta p\approx h$. Let $t$ be the time interval in which we read the length $l$. (We should note that $t$ is measured by a clock other than the suspended clock.) Then we cannot determine the force exerted by the gravitational field on the clock to a finer accuracy than $\Delta p/t$. Therefore we cannot determine the mass $m$ to a finer accuracy than $\Delta m$ given by the relation 
\begin{equation}
{{\Delta p}\over t}\approx g\Delta m. \label{eq:force}
\end{equation}

Now, according to general relativity theory, a clock, when displaced in the direction of the gravitational force by an amount $\Delta q$, changes its rate in such a way that its reading in the course of a time interval $t$ differs by an amount $\Delta \tau $ given by the relation
\begin{equation}
{{\Delta \tau }\over t}={{g\Delta q}\over {c^2}}. \label{eq:redshift}
\end{equation}
By combining (\ref{eq:force}), (\ref{eq:redshift}) and the relation  
$\Delta q \Delta p\approx h$, we see, therefore, that there is an uncertainty relation 
\begin{equation}
c^2\Delta m\Delta \tau \approx h \label{eq:time-mass-un}
\end{equation}
between the rest mass $m$ and the proper time $\tau $ of the clock.

The relativistic red-shift formula (\ref{eq:redshift}) was, of course, essential in Bohr's reasoning above. The more essential it seems to be, however, the stronger the apprehension we feel that the uncertainty relation (\ref{eq:time-mass-un}) may fail if we can think of a weighing procedure not resorting to any interaction between the clock and the  gravitational field. We check one such case in the following. 

Assume that the clock has been brought to rest after being charged with an electric charge $e$, and that a uniform electric field ${\cal E}$ is then switched on. After a short time $t$, we measure the distance  the clock has moved. (Again $t$ is the time measured by a clock other than our clock in the electric field.) Then we can know the average velocity $v$ of the clock by dividing the distance by the value of $t$, and we can determine the mass $m$ of the clock by virtue of the formula
$$e{\cal E}=m{v\over t}.$$

Assume that the determination of the distance is made with a given accuracy $\Delta q$. Then it implies a minimum latitude $\Delta p$ in the momentum of the clock, where $\Delta q\Delta p\approx h$. Hence we cannot determine the force exerted by the electric field on the clock to a finer accuracy than $\Delta p/t$. Therefore, even when the  velocity $v$ is obtained, we cannot determine the mass $m$ to a finer accuracy than $\Delta m$ given by the relation 
\begin{equation}
{{\Delta p}\over t}\approx \Delta m{v\over t}\hskip 1cm {\rm i.e.} \hskip 1cm \Delta p\approx v\Delta m. \label{eq:flat-force}
\end{equation}

Now, according to special relativity theory, when a clock has a speed $v$, its rate $\tau $ in the course of a time interval $t$ is given by the relation 
\begin{equation}
\tau =t{\sqrt {1-\left( {v\over c}\right) ^2}}. \label{eq:time-shift}
\end{equation}
On the other hand, the average velocity $v$ has an uncertainty $\Delta v$  given by the relation 
$$t\Delta v\approx \Delta q.$$
Correspondingly, the clock has an uncertainty in its rate $\tau $ of the order  $\Delta \tau $ given by 
\begin{equation}
\Delta \tau =t\cdot \Delta {\sqrt {1-\left( {v\over c}\right) ^2}}\approx {v\over {c^2}}t\Delta v\approx {v\over {c^2}}\Delta q. \label{eq:special}
\end{equation}
By combining (\ref{eq:flat-force}), (\ref{eq:special}) and the relation  
$\Delta q \Delta p\approx h$, we arrive, therefore, at the same uncertainty relation 
$$c^2\Delta m\Delta \tau \approx h$$
as (\ref{eq:time-mass-un}) obtained by Bohr's reasoning. 

Thus the uncertainty relation (\ref{eq:time-mass-un}) has been confirmed for a weighing procedure which does not rely on 
 any gravitational interaction. Moreover, in this case, the time-shift formula (\ref{eq:time-shift}) played an essential role in place of the relativistic red-shift formula.

Each of these formulae is, of course, one of the deepest and most important results in relativistic theory. The fact that these important formulae play essential roles in deriving the uncertainty relation (\ref{eq:time-mass-un}) lends some confidence as to its universality. 

The objective of this article is to show the following: The uncertainty relation (\ref{eq:time-mass-un}) can be derived satisfactorily only if we describe the system of the clock by using a Lagrangian which includes  the proper time as a dynamic variable.     

In the next section,  selecting the simplest Lagrangian which is in accord with the above approach, we examine the Hamiltonian formalism of the clock. Our conclusion is that the rest energy can be considered the momentum conjugate to the proper time. In the third section, following Dirac's procedure, we quantize the system of the clock, and we obtain the same uncertainty relation as (\ref{eq:time-mass-un}). Some comments then follow on our quantization.  

\vskip 1cm
\noindent
{\bf II. LAGRANGIAN AND HAMILTONIAN FORMALISM}

A gravitational field $g_{\mu \nu }$ and an electromagnetic field $A_{\mu }$ are assumed to be given, and we consider our clock to be one material particle moving in those fields with electric charge $e$. 

The Lagrangian which is generally used in such a case is the following:
$$L_0=-mc{\sqrt {-g_{\mu \nu }(x){\dot x}^{\mu }{\dot x}^{\nu }}}+eA_{\mu }(x){\dot x}^{\mu },$$
where $x^{\mu }\ (\mu =0, 1, 2, 3)$ are the variables and the dot denotes the differential with respect to an arbitrary parameter $\lambda $. It goes without saying that $m$ is the rest mass of the clock and that $c$ is the speed of light. 

We, however, cannot consider the proper time $\tau $ a physical quantity if we describe the system by using the Lagrangian $L_0$. On  the other hand, it is clear that the proper time of a clock is a measurable physical quantity. (It is why a clock is so named.) Hence  
we have to find another Lagrangian which is in accord with the system of the clock.
 
Our first purpose in this section is to find a Lagrangian $L$ which satisfies the following conditions:
\begin{enumerate}
\item The Lagrangian L has the proper time $\tau $ as a new variable in addition to $x^{\mu }$.  
\item The motion equations for the variables $x^{\mu }$ are invariant  between $L$ and $L_0$. 
\end{enumerate}

As a candidate we consider the Lagrangian defined by 
$$L=M\left( {\dot \tau }-{\sqrt {-g_{\mu \nu }(x){\dot x}^{\mu }{\dot x}^{\nu }}}/c\right) +eA_{\mu }(x){\dot x}^{\mu },$$
where the dynamic variables are $\tau , M$ and $x^{\mu }$. 

The Lagrange's equations of motion are as follows:
\begin{eqnarray}
&&{\dot M}=0     \label{eq:M}\\
&& {\dot \tau }={\sqrt {-g_{\mu \nu }(x){\dot x}^{\mu }{\dot x}^{\nu }}}/c  \label{eq:tau} \\
&&{d\over {d\lambda }}\left[ {M\over c}{{g_{\rho \mu }{\dot x}^{\mu }}\over {\sqrt {-g_{\mu \nu }(x){\dot x}^{\mu }{\dot x}^{\nu }}}}+eA_{\rho }(x)\right]  \nonumber \\
&&\hskip 2cm -{M\over c}{{g_{\mu \nu ,\rho }{\dot x}^{\mu }{\dot x}^{\nu }}\over {2{\sqrt {-g_{\mu \nu }{\dot x}^{\mu }{\dot x}^{\nu }}}}}-eA_{\mu , \rho }(x){\dot x}^{\mu }=0   \label{eq:third}
\end{eqnarray}

The second equation (\ref{eq:tau}) means that we can identify the variable $\tau $ with the proper time of this clock. Moreover we have $d\tau / d\lambda >0$, and therefore it is possible to change the differential with respect to $\lambda $ to one with respect to $\tau $ in the third equation (\ref{eq:third}). As a result we find that 
$${d\over {d\tau }}\left[ {M\over {c^2}}g_{\rho \mu }{\dot x}^{\mu }+eA_{\rho }(x)\right] -{M\over {2c^2}}g_{\mu \nu , \rho }{\dot x}^{\mu }{\dot x}^{\nu }-eA_{\mu , \rho }(x){\dot x}^{\mu }=0,$$
where the dot denotes the differential with respect to $\tau $. Rewriting this equation, we get 
\begin{equation}
{M\over {c^2}}\left[ {\ddot x}^{\rho }+{\Gamma ^{\rho }}_{\mu \nu }{\dot x}^{\mu }{\dot x}^{\nu }\right] =ef^{\rho \mu }{\dot x}_{\mu },  \label{eq:Motion}
\end{equation}
where ${\Gamma ^{\rho }}_{\mu \nu }$ and $f_{\mu  \nu }$ are defined by 
$${\Gamma ^{\rho }}_{\mu \nu }={1\over 2}g^{\rho \sigma }\left( -g_{\mu \nu , \sigma }+g_{\nu  \sigma , \mu }+g_{\sigma  \mu , \nu }\right) , \hskip 1cm f_{\mu \nu }=A_{\nu , \mu }-A_{\mu , \nu }. $$
On the other hand, the motion equation derived from the original Lagrangian $L_0$ is 
\begin{equation}
m\left[ {\ddot x}^{\rho }+{\Gamma ^{\rho }}_{\mu \nu }{\dot x}^{\mu }{\dot x}^{\nu }\right] =ef^{\rho \mu }{\dot x}_{\mu }.  \label{eq:o-motion}
\end{equation}
Equation (\ref{eq:Motion}) is just the same as equation (\ref{eq:o-motion}) if we identify $M$ with the constant $mc^2$. Equation (\ref{eq:M}) indicates that this identification is possible.

Thus our first purpose has been achieved. Moreover, this Lagrangian $L$ is the simplest of those which satisfy the above two conditions.

The second purpose in this section is to investigate, by using the Lagrangian $L$, the consequences of our assertion that the proper time should be considered a dynamic variable. 

We note that it is possible to propose an argument without imposing any limitation on the fields $g_{\mu \nu }$ and $A_{\mu }$. 
In such an argument, however, we have to handle the coordinate time $x^0=ct$ as a dynamic variable, and then determine certain constraint conditions for the variables. Discussion of such constraints is not essential for our purpose. We therefore assume for simplicity hereafter that the fields  $g_{\mu \nu }$ and $A_{\mu }$ are  so-called static in the following sense: 
\begin{enumerate}
\item The functions $g_{\mu \nu }$ and $A_{\mu }$ depend on only  $x^1, x^2, x^3$.
\item For $i=1, 2, 3$, we have  $g_{i 0}(=g_{0 i})=0$.
\end{enumerate}

Assuming the above conditions, we get 
$$L=M\left( {\dot \tau }-{\sqrt {f(x)^2-g_{i j}(x){\dot x}^i{\dot x}^j/c^2}}\right) +ceA_0(x)+eA_i(x){\dot x}^i,$$
where $f$ is defined by $g_{0 0}=-f^2\ (f>0)$. The dynamic variables are $\tau , M, x^i\ (i=1, 2, 3)$, and the dot denotes the differential with respect to $t$. 

The momentums conjugate to those variables are given by  
$$p_{\tau }\equiv {{\partial L}\over {\partial {\dot \tau }}}=M, \hskip 2cm p_M\equiv {{\partial L}\over {\partial {\dot M}}}=0 $$
and 
$$p_i\equiv {{\partial L}\over {\partial {\dot x}^i}}={M\over {c^2}}{{g_{i j}{\dot x}^j}\over {\sqrt {f^2-g_{j k}{\dot x}^j{\dot x}^k/c^2}}}+eA_i. $$
We have 
\begin{eqnarray*}
H_0&\equiv &p_{\tau }{\dot \tau }+p_M{\dot M}+p_i{\dot x}^i-L  \nonumber \\ 
&=&f{\sqrt {M^2+c^2g^{i j}(p_i-eA_i)(p_j-eA_j)}}-ceA_0. 
\end{eqnarray*}
If $M$ is replaced by $mc^2$, then $H_0$ is identical with the Hamiltonian which is derived from the original Lagrangian $L_0$. In our case, however, there exist two constraints:
$$\phi _1\equiv M-p_{\tau }=0, \hskip 1cm \phi _2\equiv p_M=0.$$
Taking account of these constraints, we have to consider the total Hamiltonian
$$H\equiv H_0+u_1\phi _1+u_2\phi _2,$$
where $u_1$ and $u_2$ are Lagrange's undetermined multipliers. 

The multipliers $u_1$ and $u_2$ are determined in the following manner$^3$: Poisson's bracket of $\phi _1$ and $\phi _2$ is 
$$\{ \phi _1, \phi _2\} = 1$$
and therefore we have 
$${\dot \phi }_1=\{ \phi _1, H\} \approx u_2, $$
$${\dot \phi }_2=\{ \phi _2, H\} \approx  -u_1-{{fM}\over {\sqrt {M^2+c^2g^{i j}(p_i-eA_i)(p_j-eA_j)}}}, $$
where the symbol \lq \lq $\approx $'' denotes the weak equality defined by the constraints $\phi _1= \phi _2=0$. Hence, the consistency conditions
$${\dot \phi }_1\approx 0 \hskip 1cm {\rm and}\hskip 1cm {\dot \phi }_2\approx 0$$
require the multipliers $u_1$ and $u_2$ to be 
$$u_1=-{{fM}\over {\sqrt {M^2+c^2g^{i j}(p_i-eA_i)(p_j-eA_j)}}} \hskip 5mm {\rm and} \hskip 5mm u_2=0, $$
which give 
\begin{equation}
H=H_0-{{fM(M-p_{\tau })}\over {\sqrt {M^2+c^2g^{i j}(p_i-eA_i)(p_j-eA_j)}}}.    \label{eq:Hamiltonian}
\end{equation}

Hamilton's canonical equations of motion are as follows:
\begin{eqnarray*}
&&{\dot \tau }={{\partial H}\over {\partial p_{\tau }}}={{fM}\over {\sqrt {M^2+c^2g^{i j}(p_i-eA_i)(p_j-eA_j)}}} \cr
&&{\dot p}_{\tau }=-{{\partial H}\over {\partial \tau }}=0\cr
&&{\dot M}={{\partial H}\over {\partial p_M}}=0\cr
&&{\dot p}_M=-{{\partial H}\over {\partial M}}\approx 0\cr
&&{\dot x}^i={{\partial H}\over {\partial p_i}}\approx {{\partial H_0}\over {\partial p_i}}\cr
&&{\dot p}_i=-{{\partial H}\over {\partial x^i}}\approx -{{\partial H_0}\over {\partial x^i}}\cr
\end{eqnarray*}

Defining a matrix $W_{i j}$ by 
$$W_{i j}\equiv \{ \phi _i, \phi _j\} =\pmatrix{ 0 & 1 \cr -1 & 0 \cr },$$
we can write Dirac's bracket:
\begin{eqnarray*}
\{ A, B\} _D&=&\{ A, B\} -\sum _{i, j=1}^2\{ A, \phi _i\} W^{-1}_{i j}\{ \phi _j, B\} \cr
&=&\{ A, B\} +\{ A, \phi _1\} \{ \phi _2, B\} -\{ A, \phi _2\} \{ \phi _1, B\} .\cr
\end{eqnarray*}
We can easily calculate Dirac's brackets between the canonical variables:
$$\{ \tau , p_{\tau }\} _D=\{ \tau , M\} _D=1, \hskip 1 cm \{ x^i, p_j\} _D={\delta ^i}_j, \hskip 1 cm {\rm the\ \ others}=0.$$

We are now in a position to be able to state our conclusions in this section.

It is easily shown that  
$$\phi _1, \hskip 5 mm \phi _2, \hskip 5mmT\equiv \tau - p_M, \hskip 5mmE\equiv p_{\tau }, \hskip 5mm x^i, \hskip 5mm p_i, \hskip 5 mm  (i=1, 2, 3)$$
are canonical variables, and therefore the variables $T, E, x^i, p_i (i=1, 2, 3)$ can be interpreted as canonical variables on the submanifold defined by the constraints $\phi _1=\phi _2=0$. We can show also that 
$$\{ A, B\} _D={{\partial A}\over {\partial T}}{{\partial B}\over {\partial E}}-{{\partial A}\over {\partial E}}{{\partial B}\over {\partial T}}+\sum _{i=1}^3\left( {{\partial A}\over {\partial x^i}}{{\partial B}\over {\partial p_i}}-{{\partial A}\over {\partial p_i}}{{\partial B}\over {\partial x^i}}\right)  $$
on the submanifold.

Since we have that  
$$T=\tau  \hskip 1cm {\rm and}\hskip 1cm  E=M(=mc^2) $$
on the submanifold defined by $\phi _1=\phi _2=0$, it follows from the above that the rest energy $mc^2$ is considered the momentum conjugate to the proper time $\tau $. 

\vskip 1cm 
\noindent
{\bf III. QUANTIZATION AND DISCUSSIONS}

Thus we have arrived at the following conclusion: If we accept the 
view that we should describe a clock by using a Lagrangian which includes the proper time as a dynamic variable like the positions $x^i$, then we find that the rest energy $E=mc^2$ turns out to be the general momentum conjugate to the proper time, and that $\tau , E, x^i$ and $p_i$ are canonical variables of the system.

Since $\tau , E, x^i, p_i$ are the canonical variables, if we quantize the system by Dirac's procedure, there are corresponding operators 
$${\hat \tau }, \hskip 1cm {\hat E}, \hskip 1cm {\hat x}^i, \hskip 1cm {\hat p}_i\hskip 1cm (i=1, 2, 3) $$
which satisfy the commutation relations 
\begin{equation}
[{\hat \tau }, {\hat E}]=[{\hat x}^i, {\hat p}_i]=i{\hbar }. \label{eq:commutation}
\end{equation} 
The relation $[{\hat \tau }, {\hat E}]=i{\hbar }$ in (\ref{eq:commutation}) leads us to the uncertainty relation  
\begin{equation}
c^2\Delta m\Delta \tau \geq {\hbar }/2 \label{eq:uncertainty}
\end{equation}
which was argued in the Introduction to this article.

Our quantization leads to some desirable results besides the uncertainty relation (\ref{eq:uncertainty}), but at the same time gives rise 
 to some problems.  

First, we should make some comment on the problems. In our quantization, the operators ${\hat \tau }, {\hat E}, {\hat x}^i$ and ${\hat p}_i \hskip 2mm (i=1, 2, 3)$ can be represented in the Hilbert space composed of square integrable functions of $\tau , x^1, x^2$ and $x^3$. In particular, the operator ${\hat E}$ is represented by the differential operator $-i{\hbar }\partial /\partial \tau $, and therefore the rest energy ${\hat E}$ cannot have any discrete spectrum. Furthermore, this Hilbert space includes some states in which the mean values of ${\hat E}$ are negative. 

The problems of the continuous mass spectrum and of the negative mass are inevitable in our formulation. The authors cannot, at present, judge whether these characteristics are desirable or not. These problems will be discussed in a subsequent paper from a rather different viewpoint.      

Secondly, we focus our attention on some positive aspects of our quantization. We restrict ourselves, for simplicity, to the case in which the space-time metric is flat and $A_{\mu }=0$. Then the Hamiltonian in (\ref{eq:Hamiltonian}) is rather simple and the Hamiltonian operator has the form   
$${\hat H}\equiv \sqrt {{\hat E}^2+c^2{\bf {\hat p}}^2}.$$
(We omit, hereafter, the hats representing the operators since there is no possibility of misunderstanding.)  

For the Heisenberg representation of the operator $\tau $
$$\tau (t)=e^{itH/{\hbar }}\tau e^{-itH/{\hbar }},$$
we find that  
\begin{equation}
{d\over {dt}}\tau (t)={i\over {\hbar }}e^{itH/{\hbar }}[H, \tau ]e^{-itH/{\hbar }}={E\over {\sqrt {E^2+c^2{\bf p}^2}}} \label{eq:Lorentz}
\end{equation}
by virtue of 
$$[\tau , H]=i{\hbar }{E\over {\sqrt {E^2+c^2{\bf p}^2}}}.$$
Hence we  have
\begin{equation}
\tau (t)={E\over {\sqrt {E^2+c^2{\bf p}^2}}}t+\tau . \label{eq:first}
\end{equation}
We note that the last term of (\ref{eq:Lorentz}) is the operator which represents the time delay of the moving clock. 

We can moreover show that
\begin{eqnarray*}
{d\over {dt}}\tau (t)^2&=&{E\over {\sqrt {E^2+c^2{\bf p}^2}}}\tau (t)+\tau (t){E\over {\sqrt {E^2+c^2{\bf p}^2}}}  \nonumber \\
&=&2{{E^2}\over {E^2+c^2{\bf p}^2}}t+\left[ {E\over {\sqrt {E^2+c^2{\bf p}^2}}}, \ \tau  \right] _+,  
\end{eqnarray*}
where we have used equation (\ref{eq:first}), and where $[ A, \ B]_+$  denotes the anti-commutator of operators $A$ and $B$. Integrating this, we have 
$$\tau (t)^2={{E^2}\over {E^2+c^2{\bf p}^2}}t^2+\left[ {E\over {\sqrt {E^2+c^2{\bf p}^2}}}, \ \tau \right] _{+}t+\tau ^2. $$
Hence the standard deviation $\Delta \tau (t)$ in a state $\psi $ is given by
\begin{eqnarray}
(\Delta \tau (t))^2&\equiv &\langle \tau (t)^2\rangle -\langle \tau (t)\rangle ^2 \nonumber \\
&=&\left( \langle  {{E^2}\over {E^2+c^2{\bf p}^2}}\rangle -\langle  {E\over {\sqrt {E^2+c^2{\bf p}^2}}}\rangle ^2\right) t^2 \nonumber \\
&&\hskip 1cm +\left( \langle \left[ {E\over {\sqrt {E^2+c^2{\bf p}^2}}}, \ \tau \right] _{+}\rangle -2\langle  {E\over {\sqrt {E^2+c^2{\bf p}^2}}}\rangle \langle \tau \rangle \right) t \nonumber \\
&&\hskip 4cm +\left( \langle \tau ^2\rangle -\langle \tau \rangle ^2\right) , \label{eq:deviation}
\end{eqnarray}
where $\langle A\rangle $ denotes the mean value of an operator $A$ in the state $\psi$. 

Here we must introduce some approximations: We assume that the Hamiltonian operator has a very sharp value (say $\cal E$) in the state $\psi $. This assumption seems to be natural since the clock is moving as a free particle. Under this assumption, we can approximately estimate the two terms in (\ref{eq:deviation}) in the following manner;
$$\langle  {{E^2}\over {E^2+c^2{\bf p}^2}}\rangle -\langle  {E\over {\sqrt {E^2+c^2{\bf p}^2}}}\rangle ^2\approx {1\over {{\cal E}^2}}\left( \langle E^2\rangle -\langle E\rangle ^2\right) , $$
\begin{equation}
\langle \left[ {E\over {\sqrt {E^2+c^2{\bf p}^2}}}, \ \tau \right] _{+}\rangle -2\langle  {E\over {\sqrt {E^2+c^2{\bf p}^2}}}\rangle \langle \tau \rangle \approx {1\over {\cal E}}\left( \langle [E, \ \tau ]_{+}\rangle -2\langle E\rangle \langle \tau \rangle \right) .\label{eq:vanish}
\end{equation}
On the other hand, the term $\langle [E, \ \tau ]_{+}\rangle -2\langle E\rangle \langle \tau \rangle $ in (\ref{eq:vanish}) often vanishes, as it does in the case of all optimal simultaneous measurements of $E$ and $\tau $. (We can easily check it by setting, for example, $\tau =i{\hbar }\partial /\partial E$ and $\psi =$ a Gaussian function of $E$.) Taking this cancellation into account, we neglect the second term in (\ref{eq:deviation}).

Thus we have arrived at 
$$(\Delta \tau (t))^2\approx {1\over {{\cal E}^2}}(\Delta E)^2t^2+(\Delta \tau )^2,$$
and, by virtue of the inequality 
$${1\over {{\cal E}^2}}(\Delta E)^2t^2+(\Delta \tau )^2\geq  {2\over {\cal E}}\Delta \tau \Delta E t, $$
we finally have
\begin{equation}
(\Delta \tau (t))^2\geq {{\hbar }\over {\cal E}}t, \label{eq:final} 
\end{equation}
where we have used the uncertainty relation $\Delta \tau \Delta E\geq {\hbar }/2$  of  (\ref{eq:uncertainty}). 

When the motion of the clock is so slow that the value of ${\cal E}$ is approximately equal to $mc^2$, then our inequality (\ref{eq:final}) has the form 
\begin{equation}
(\Delta \tau (t))^2\geq {{\hbar }\over {mc^2}}t, \label{eq:final-nonrela}
\end{equation}
which exactly coincides with an inequality derived by Salecker and Wigner from another point of view (see  Eq. (6) in Ref. 4).  

In conclusion, we should make some comment on the meaning of our results to physics. 

Bohr and Rosenfeld stressed the principle that every proper theory should provide in and by itself its own means for defining the quantities with which it deals. One of the key points this principle makes is that we should analyze the means of measuring those quantities in order to argue the consistency of a physical theory. In their case, they succeeded in showing that the definition of the standard quantization of electromagnetic field is consistent in the above sense by discussing the means of measuring the classical electromagnetic field$^{5,6}$.

Several authors have applied this principle to the theory of relativity to find a consistent quantization of the space-time geometry. The theory  deals with such quantities as the metric tensor, the curvature tensor, the covariant derivative and connection coefficients. The measurement of the distance between two events is most fundamental in the procedures by which we measure these quantities$^{}$. For this we require 
the concept of a clock$^{7,10}$, and the clock cannot be independent of the various physical laws. Thus, if the above principle should be a general feature of physical theory, a consistent formulation of the quantization of the space-time geometry should have some inherent relation with various limitations on the accuracy of the clock resulting from the physical laws.

Various gedanken experiments on such limitations have been proposed and elaborated on for some fifty years$^{4,7-16}$. In many of them, however, the clock is assumed to have some structure, from which starting point the argument is developed. It seems uncertain therefore
 whether their results are  universal or not. Moreover, different studies sometimes reach different conclusions. Our objective in the present paper was to propose an attempt to dispose of this ambiguity.
We showed the following : (a) There is an uncertainty relation between the proper time and the rest mass of a clock independent of its structure (see Eq. (3)). (b) A limitation on the accuracy of the clock is derived from the uncertainty relation in a natural way (see Eqs. (19) and (20)).

The subject raised here has been argued, despite its importance, only at the level of thought experiments. The authors are uneasy with this situation, and think that the time has come to argue it at a more positive level. We hope that the importance of this subject is recognized and that, for example, the relation (20) is verified by experiment in the near future.

\null
\vskip 1cm
\noindent
$^1$A.Pais, {\it \lq Subtle is the Lord ...' The Science and the Life of Albert Einstein} (Oxford University, 1982).\hfill \break 
$^2$M.Jammer, {\it The philosophy of quantum mechanics}(John Wiley \& Sons, Inc., 1974).\hfill \break 
$^3$P.A.M.Dirac, Canad. J. Math. {\bf 2}, 129(1950); Proc. Roy. Soc. (London) {\bf A 246}, 326(1958). \hfill \break 
$^4$H.Salecker and E.P.Wigner, Phys. Rev. {\bf 109}, 571(1958).
\hfill \break 
$^{5}$N.Bohr and L.Rosenfeld, Mat.-Fys. Medd. Dan. Vid. Selsk. {\bf 12}, no.8 (1933); Phys. Rev. {\bf 78}, 794 (1950).\hfill \break 
$^{6}$L.D.Landau and R.Peierls, Z. Phys. {\bf 69}, 56 (1931);  in {\it Collected Papers of Landau}, ed. D.ter Haar, (Gordon and Breach, New York, 1965), pp. 40-51.    \hfill \break 
$^{7}$E.P.Wigner, Rev. Mod. Phys. {\bf 29}, 255 (1957).\hfill \break 
$^{8}$A.Peres and N.Rosen, Phys. Rev. {\bf 118}, 335 (1960).\hfill \break 
$^{9}$C.A.Mead, Phys. Rev. {\bf 135}, B849 (1964).\hfill \break 
$^{10}$R.F.Marzke and J.A.Wheeler, in {\it Gravitation and Relativity}, eds. H.Y.Chiu and W.F.Hoffman,  (W.A.Benjamin, New York, 1964).\hfill \break 
$^{11}$F.K{\'a}rolyh{\'a}zy, A.Frenkel, and B.Luk{\'a}cs,  in {\it Quantum Concepts in Space and Time}, eds. R.Penrose and C.J.Isham
, (Clarendon, Oxford, 1986).\hfill \break 
$^{12}$A.Charlesby, Radiat. Phys. Chem. {\bf 33}, 487 (1989).\hfill \break 
$^{13}$L.Di{\'o}si and B.Luk{\'a}cs, Phys. Lett. A {\bf 142}, 331 (1989).\hfill \break 
$^{14}$F.K{\'a}rolyh{\'a}zy, in {\it Sixty-Two Years of Uncertainty}
, ed. A.I.Miller, (Plenum, New York, 1990).\hfill \break 
$^{15}$M.Maggiore, Phys. Lett. B {\bf 304}, 65 (1993).\hfill \break 
$^{16}$S.Doplicher, Ann. Inst. Henri Poincare Phys. Theor. {\bf 64}, 543 (1996).

\vfill
\end{document}